\documentclass[twocolumn,showpacs,preprintnumbers,amsmath,amssymb]{revtex4}
\usepackage{graphicx}
\usepackage{dcolumn}
\usepackage{bm}
\usepackage{footnote}

\newcommand \beq{\begin{eqnarray}}
\newcommand \eeq{\end{eqnarray}}
\newcommand \bea{\begin{eqnarray}}
\newcommand \eea{\end{eqnarray}}

\newcommand \bp{{\mathbf p}}

\newcommand \bq{{\mathbf q}}

\usepackage{amsmath}
\usepackage{amssymb}

\def\simle{\mathrel{
     \rlap{\raise 0.511ex \hbox{$<$}}{\lower 0.511ex \hbox{$\sim$}}}}

\begin{document}
\title{Entropy in the quark-hadron transition}
\author{
Kanako Yamazaki$^a$,\ T. Matsui$^a$,\ and Gordon Baym$^{bc}$ 
}
\affiliation{$^a$Institute of Physics, University of Tokyo, Komaba, Tokyo, Japan}
\affiliation{$^b$Department of Physics,  University of Illinois, 
Urbana, IL, 61801 USA}

\begin{abstract}

   We study, in the PNJL model, how the entropy of interacting quarks reflects the change in the effective degrees of freedom as the temperature increases through the quark-hadron phase transition.  With inclusion of mesonic correlations, the effective degrees of freedom change from those of pi and sigma mesons at low temperatures to those of free quarks at high temperatures, with a resultant second order phase deconfinement transition in the chiral limit.


\end{abstract}
\pacs{12.38.-t,12.38.Mh, 11.10.Wx, 25.75.Nq}
\maketitle

\section{Introduction} 
An intriguing and as yet unsolved question in modern nuclear physics is how the effective degrees of freedom change from hadronic at low temperature to those of quarks and gluons at high temperature \cite{Wein81}. 
A useful probe of the  nature of the deconfinement phase transition
 is the entropy, since it reflects how the effective degrees of freedom change.
The entropy plays a key role in the evolution of the high temperature quark-gluon plasma in ultrarelativistic nucleus-nucleus collisions at RHIC and LHC.  Its importance in high energy collisions was first recognized by Landau in his hydrodynamic model of multi-particle production \cite{Landau:1953gs}, and has been reemphasized in early studies of the hydrodynamic evolution of the quark-gluon plasma in ultrarelativistic nucleus-nucleus collisions \cite{Bjo83,Baym83,Baym83a,Kaj83,Gyu84}. 
The entropy of the quark-gluon plasmas has been computed by perturbative QCD at finite temperatures in the weak coupling regime, where  quasiparticle modes carry the entropy \cite{Blaizot:1999ip}. 
It has also been measured by Monte Carlo lattice gauge theory in the strong coupling regime \cite{Borsanyi:2013cga}. 

Here we compute the entropy using the Polyakov-Nambu-Jona-Lasinio 
model for interacting quarks at zero net baryon density. 
The model was originally designed \cite{Fukushima:2003fw} to describe both the chiral phase transition and the deconfining transition, with an NJL-type \cite{Nambu:1961tp}  effective quark interaction respecting chiral symmetry \cite{HK94}; the model is supplemented by the Polyakov loop, which plays the role of an order parameter for deconfinement \cite{Polyakov:1978vu}.
The model has been reformulated as a mean field theory with a background uniform temporal color gauge field, which acts on quarks 
as a color-dependent imaginary chemical potential \cite{Megias:2004hj, Ratti:2005jh, Ratti:2007jf}. 
In the confining phase, the destructive interference of the (complex) distributions of the three differently colored quarks converts the quark quasiparticle distribution effectively into a distribution of ``quark triads," a color singlet combination of three massive quark quasiparticles each of the same momentum.  In the high temperature deconfining phase, the vanishing background gauge field brings about a free nearly massless quark distribution.   

One must, in determining the entropy, also take into account hadronic excitations. These  
can be included by computing the mesonic correlation energy \cite{Rossner:2007ik, Blaschke:2007np, Hansen:2006ee,  Wergieluk:2012gd, Benic:2013eqa}, which  indeed reproduces, at zero baryon chemical potential, the
hadron gas pressure at low temperatures \cite{Yamazaki:2012ux, Yamazaki:2013yua}.
The pressure of colored quarks is depleted by phase cancellations between the distributions of color-triplet quarks, implying
that the entropy of the low temperature, zero baryon chemical potential phase
is carried by pionic excitations -- Nambu-Goldstone (NG) modes associated with spontaneously broken chiral symmetry.  
As chiral symmetry is restored, above a certain temperature, these NG bosons disappear and
the carriers of the entropy change.  While chiral symmetry restoration may in principle lead to a discontinuous change of the entropy across the transition, as a simple bag model calculation gives \cite{Baym76}, 
we show here that the entropy changes continuously with temperature as the collective meson modes cease to exist as isolated poles and dive into the continuum of pair excitations of quark quasiparticles
\footnote{ We note that this mechanism of meson dissolution resembles, but is not the same as, that of the disappearance of bound states of a heavy quark and its antiparticle caused by screening of binding potential \cite{MS86}.}.
The entropy carried by the collective mesonic modes is transferred to the non-collective pair excitations of quark quasiparticles on the higher temperature side. 

\section{Mean field quark entropy with color correlations} 
We begin with the mean field calculation of the pressure, including color correlations between quarks through Polyakov loops, and then include mesonic correlations, calculating the entropy from
$s (T) = \partial p/\partial T$ at fixed chemical potentials.
In the first, mean field, approximation the pressure of the PNJL model is given at zero baryon chemical potential as a function of the temperature, the effective quark mass $M$, and the Polyakov loop parameter $\Phi$,  by
\begin{eqnarray}
p_{\rm MF}(T, M, \Phi ) &=& p_{\rm MF}^{\rm ren}(M) \\
&+& \gamma_q \int \frac{d^3p}{(2\pi )^3} \frac{p^2}{3E_p } {\rm tr}_c  f (E_p, A_0) - \mathcal{U}(T, \Phi )\nonumber .  \label{pMF}
\label{pmf}
\end{eqnarray}
The first term is the  cutoff-regularized vacuum pressure of the quarks in the filled Dirac sea,
\begin{equation}
p_{\rm MF}^{\rm ren}(M) = - \frac{1}{4G} (M - m_0)^2 + \frac{\gamma_q}{2} \int^{\Lambda} \frac{d^3p}{(2\pi )^3} E_p - \Delta P_{\rm vac},
\label{pmean}
\end{equation}
in which the first term
is the condensate pressure due to the shift of the effective quark mass $M$ from its bare value $m_0$, and $-G$ is the NJL coupling constant; the second term is 
the pressure of the negative energy quark sea, with $\Lambda$ the ultraviolet cutoff; $\gamma_q = 2 \times 2 \times 3\times N_f $ counts spin, particle-antiparticle, color, and flavor degeneracy,  and $E_p=\sqrt{p^2+M^2}$ is the quark-quasiparticle energy; 
the renormalization constant $\Delta P_{\rm vac}$ is chosen so that the pressure vanishes at zero temperature.  
The second term in Eq.~(\ref{pmf}) is the thermal quark pressure, where
\begin{eqnarray}
f (E_p, A_0 ) = \frac{1}{e^{\beta (E_p + i A_0)} +1}  , 
\label{fA0}
\end{eqnarray}
is the quark distribution function; the trace is over quark colors, and $A_0$ is the background temporal color gauge field, specified below.
The final term in (1) is the gluon pressure due to the effective potential $\mathcal{U}(T, \Phi )$ for $A_0$, where the Polyakov loop parameter $\Phi$ is defined in terms of $A_0$ below.

  In general the background gauge field can be chosen in the diagonal form $A_0 = \varphi_3 \lambda_3 + \varphi_8 \lambda_8$, where 
the $\lambda$'s are usual SU(3) matrices.   Extremization of the pressure with respect to the $\varphi$'s builds in color neutrality of the
system.   For simplicity here we include only the $\lambda_3$ term and write 
$\beta A_0 = {\rm diag} ( \phi_1, \phi_2, \phi_3 ) =  {\rm diag} ( \varphi, -\varphi, 0 )$.
The trace is the sum over the angles $\phi_i$ for the three color components 
of quark fields.  The quark mean field pressure is then,   
\begin{eqnarray}
p_{\rm MF}^{\rm q} (T, M, \phi ) &=& 
\gamma_q \int \frac{d^3p}{(2\pi )^3} \frac{p^2}{3E_p } \sum_i f (\beta E_p  + i \phi_i). 
\nonumber   
\label{pquark}
\end{eqnarray}
The Polyakov loop parameter $\Phi$ is defined by 
\begin{equation}
\Phi = \frac{1}{3} \langle {\cal P} e^{ i \int_0^\beta d \tau A_0} \rangle,
\end{equation}
and is thus related to the angle $\varphi$ by 
\begin{equation}
\Phi  = \bar{\Phi } = \frac13 (1+2 \cos \varphi).
\end{equation}
Thus $\Phi$ varies from 1 to 0 as $\varphi$ varies from 0 to $2\pi/3$; $\Phi = 1$ describes the fully deconfined phase, and $\Phi=0$ describes the fully confined phase.  Because of the explicit $i$'s, 
 the quark distribution function  cannot be readily interpreted as an average distribution in phase space; however,  in the sum over the three quark colors,
 the imaginary parts cancel and we find the real quenched quark distribution
 \cite{Fukushima:2003fw},
\begin{eqnarray}
f_{\Phi }(E_p) & = &\frac13 \sum_i f (\beta E_p  + i \phi_i) \nonumber \\
& = & \frac{\bar{\Phi }e^{2\beta E_p}+2 \Phi e^{\beta E_p}+1}{e^{3\beta E_p}+3\bar{\Phi }e^{2\beta E_p}+3\Phi e^{\beta E_p}+1}.
\label{quenched}
\end{eqnarray}
The effective gluon free energy can be modeled in terms of $\Phi$ by
\begin{eqnarray}
\frac{\mathcal{U}(T, \Phi )}{T^4}=-\frac{1}{2}b_2(T)\bar{\Phi }\Phi -\frac{1}{6}b_3(\Phi ^3+\bar{\Phi }^3)
+\frac{1}{4}b_4(\bar{\Phi }\Phi )^2 \nonumber \\
\eeq
where
\beq
b_2(T)&=&a_0+a_1\left( \frac{T_0}{T}\right)+a_2\left( \frac{T_0}{T}\right) ^2
+a_3\left( \frac{T_0}{T}\right) ^3;
\end{eqnarray}  
the parameters are chosen 
so that the value of $\Phi$ that minimizes $\mathcal{U}$ vanishes at zero temperature, and approaches unity at high temperatures, where $- \mathcal{U}(T, \Phi )$ approaches the Stephan-Boltzmann free gluon gas pressure \cite{Yamazaki:2012ux}.

The effective quark mass $M$ and the Polyakov loop $\Phi$ are then determined by maximizing the pressure (or equivalently minimizing the free energy). 
Taking the extremum of $p_{\rm MF}$ with respect to $M$, we obtain the gap equation: 
\begin{equation}
M - m_0 = 2 G \langle {\bar q} q \rangle.
\label{gap}
\end{equation}  
The quark scalar condensate 
is given by 
\beq
\langle {\bar q} q \rangle  
=  \gamma_q \int^{\Lambda} \frac{d^3p}{(2\pi )^3} \frac{M}{E_p} \left( \frac{1}{2} -  f_{\Phi }(E_p) \right) .
  \label{tad}
\eeq
with the cutoff $\Lambda$ regularizes the otherwise divergent momentum integral. 
The temperature dependence of $M$ is weakened in the confining phase, compared to that of free quarks, since here only color singlet thermal quark excitations are allowed. 

After extremizing $P$ with respect to $M$ and $\phi$, we obtain the entropy,
\begin{eqnarray}
s_{MF}(T, \Phi ) 
& = & 3\gamma_q \int \frac{d^3p}{(2\pi )^3}
\frac{p^2}{3E_p}\frac{\partial f_{\Phi }(E_p)}{\partial T}-\frac{\partial \mathcal{U}(T, \Phi )}{\partial T},
\nonumber \\
\label{Smf}
\end{eqnarray} 
where the partial derivative can be taken at fixed $M$ and $\Phi$.
The first term is the entropy carried by the quark quasiparticle excitations, and the second is effectively the entropy carried by gluon excitations.  

  The total entropy can be written in  terms of the
individual quark distribution functions as
\beq
   s_{MF} && =
   \nonumber\\&& -\gamma_q \int \frac{d^3p}{(2\pi )^3}\sum_{i} \left(f_{pi}\ln f_{pi} + (1-f_{pi})\ln(1-f_{pi})\right) \nonumber\\
   & &+ \frac{\varphi}{T}\frac{\partial U}{\partial \varphi}  - \frac{\partial U}{\partial T} .
 \eeq 
 The $U$ terms represent the entropy of the gluons, plus the extra quark entropy owing to
 the lack of color neutrality in the quark sector alone.  
The total entropy obeys the usual thermodynamic relation at zero chemical potential,
\begin{equation}
Ts = P + \epsilon  
\end{equation}
where $\epsilon = 3\gamma_q \int_pE_p f_\Phi+U-T\partial U/\partial T$ is the total energy density.

   Color correlations among quarks generally reduce the quark entropy from that of a Fermi gas of quarks and antiquarks.  
In the confining limit, $\Phi = 0$, where
 the quenched quark distribution turns into that of a quark triad: 
\begin{equation}
f_{\Phi = 0 }(E_p) = \frac{1}{e^{3\beta E_p}+1}, 
\end{equation}
the entropy of the quarks, as well as the anti-quarks,  is maximally reduced and assumes the canonical form,  
$$ -  \frac12\gamma_q \sum_p \left[f_{\Phi = 0 }\ln f_{\Phi = 0 } + (1-f_{\Phi = 0 })\ln(1-f_{\Phi = 0 })\right]$$ for the independent quark triad excitations;  the $\gamma_q/2$, the degeneracy factor for quarks alone, is the same as  that of light quarks, although the quarks are now massive excitations due to mean field and color correlations.
While it is tempting to interpret the entropy of quark triads as that carried by baryons of mass $3 M$ by recasting $3p \to p_B$ and $3E_p \to E_B$, the mismatch of the degrees of freedom makes the entropy from triads smaller than the entropy expected from baryons by a factor $3^{-3} = 1/27$ \cite{Yamazaki:2012ux}.  The point is that the quark triad is not really a composite baryon consisting of three quarks, but is more like a color-averaged single quark carrying the same momentum and spin quantum numbers as the original quark.  

\section{mesonic correlation entropy}
The total entropy is a sum of the mean field quark entropy, $s_{MF}$, which changes continuously with temperature through the transition region, and  the entropy of collective mesonic-like excitations at low temperature.  To include the latter excitations we need to go one step beyond mean field.  The low energy excitations we consider here are pion-like and sigma-like.   For simplicity we refer to such excitations as ``mesons" in the following.

  In the hadronic phase, pions, which are massless in the chiral limit ($m_0 = 0$), and sigma mesons contribute a pressure,
\begin{equation}
p_{\rm meson} (T) = \sum_\nu\gamma_\nu  \int \frac{d^3 q}{(2\pi )^3} \frac{q^2}{3 \omega_{q}} \left(\frac12  +   f_B (\omega_q) \right), 
\label{pmeson2}
\end{equation}
where the index $\nu$ denotes the meson type;  $\gamma =3$, $\omega_q= q$ for pions, and $\gamma = 1$, $\omega_q= \sqrt{q^2 + m_\sigma^2}$ for sigma mesons.  The $\frac12$ term is the divergent vacuum pressure, which is removed by renormalization. 
The entropy of the meson gas, $s_{\rm meson} (T) = \partial p_{\rm meson} (T)/\partial T$, has the canonical form  
\begin{eqnarray}
s_{\rm meson} (T) 
& = & \sum_\nu\gamma_\nu  \int \frac{d^3 q}{(2\pi )^3} \sigma_{B\nu}(\omega_q),
\label{smeson}
\end{eqnarray}
where
\begin{eqnarray}
\sigma_B (\omega) =  (1 +  f_B (\omega) ) \ln  (1 +  f_B (\omega) )- f_B (\omega) \ln  f_B (\omega),\nonumber\\
\end{eqnarray}
with $f_B (\omega) = 1/(e^{\beta \omega} -1 )$ the bosonic distribution function.

  In general, the mesonic modes appear as collective modes of the quarks.
We derive the mesonic entropy from the mesonic correlation pressure,  $p_{corr} \equiv p(G) - p_{\rm MF}^{\rm ren}(G)$, where $p(G)$ is the full pressure
at coupling constant $G$.   The correlation pressure is given, in the random phase approximation, in terms of Matsubara frequencies by
\begin{eqnarray}
  p_{\rm corr} (T) = \frac{1}{2\beta}\sum_{\nu}\sum_{\bq, n} \ln\left(1-2G\Pi_\nu(\omega_n,q)\right),
  \label{pcorr0}
\end{eqnarray}
where the index $\nu$ runs over the four pion and sigma degrees of freedom.  Here
$\Pi_\nu (\omega, q)$ is the quark polarization of the mean field distribution: 
\begin{eqnarray}
& & \Pi_\nu  (\omega_n, q)  =  - \frac{1}{\beta}\int \frac{d^3 p}{(2\pi)^3}  \nonumber \\
& & \times \sum_{m,i} {\rm Tr} \left[ \Gamma_\nu S_i (\varepsilon_m + \omega_n, p + q) \Gamma_\nu S_i (\varepsilon_m, p) \right],
\label{polarization1}
\end{eqnarray}
where
 $S_i (\varepsilon, p) = [(\varepsilon - i \phi_i T) \gamma_0 -  \bp \cdot {\boldsymbol \gamma} - M ]^{-1}$ is the quark quasiparticle  propagator.
The trace is over Dirac as well as flavor indices and the sum is over the quark color index $i$.   The sum over Matsubara frequencies 
yields
\begin{eqnarray}\Pi_\nu  (\omega, q) & =&  - \int \frac{d^3 p}{(2\pi)^3} \int_C
\frac{d \varepsilon}{2 \pi i}  f_F (\varepsilon)
\nonumber \\
& & \qquad \times \sum_i {\rm Tr} \left[ \Gamma_\nu S_i (\varepsilon + \omega, p + q) \Gamma_\nu S_i (\varepsilon, p) \right],
\nonumber \\
\label{polarization}
\end{eqnarray}
where $\Gamma_\pi = i \gamma_5 \tau_\nu$,  $\Gamma_{\sigma} = 1$, and
$f_F ( \varepsilon) = 1/(e^{\beta \varepsilon} +1)$. 

  The correlation pressure (\ref{pcorr0}) can be derived in several equivalent ways.  Reference \cite{Yamazaki:2012ux} 
evaluated the path integral of the effective mesonic action, obtained by integrating out the Grassmann quark fields variables, over the remaining auxiliary mesonic fields, making a Gaussian approximation around the saddle point and neglecting meson-meson interactions.  This result can be equivalently seen by differentiating the total pressure
 $P = (T/V) \ln {\rm Tr} e^{- \beta {\hat H}}$, with respect to the coupling constant $G$.  Schematically,  with ${\hat H} = {\hat H}_0 -G  \int d^3x q (x) {\bar q} (x)  {\hat \tau} q (x) 
  $  in terms of Dirac quark field operators $q (x)$,
 \begin{equation}
\frac{\partial P}{\partial G} = \frac{1}{V}\int d^3 x \langle {\bar q} (x)  q (x) {\bar q} (x) q (x) \rangle,
\label{PG}
\end{equation}
where, for simplicity we focus on the scalar field.
(Including the pseudoscalar interaction in the NJL model, $ - G \int d^3 x {\bar q} (x)  \gamma_5 {\hat \tau} q (x) {\bar q} (x)  \gamma_5 {\hat \tau} q (x) $, the right side of (\ref{PG}) acquires an additional term, $- \langle {\bar q} (x)  \gamma_5 {\hat \tau} q (x) {\bar q} (x)  \gamma_5 {\hat \tau} q (x) \rangle $.)
The right side of Eq.~(\ref{PG}) can be expressed, in the presence of a uniform scalar condensate 
$\langle {\bar q} q \rangle \neq 0$,  as the sum of the condensate pressure and the pressure due to scalar density fluctuations.  In terms of the Fourier components of the scalar density propagator,
\beq
D_s (x,\tau) 
 = -i \langle T\left( n_s (x,  \tau ) n_s (0, 0) \right \rangle,
\eeq
where
$ n_s (x, \tau) = e^{{\hat H}\tau} {\bar q} (x) q (x) e^{-{\hat H}\tau} - \langle {\bar q} q \rangle$, 
one has 
\begin{eqnarray}
 \frac{1}{V}\int d^3 x \langle {\bar q} (x)  q (x) {\bar q} (x) q (x) \rangle  = &
\nonumber \\
  \langle {\bar q} q \rangle^2 -& \frac{1}{\beta}\sum_{\bq, n} D_s ( \omega_n, q).
\label{Dint}
\end{eqnarray}
Calculating  the scalar density propagator in the random phase approximation by summing the Dyson series with the quark polarization
taken to be the lowest order scalar density fluctuation, we have
\begin{equation}
D_s (\omega_n, q) = \frac{ \Pi_\sigma  (\omega_n, q) }{1 - 2 G \Pi_\sigma (\omega_n, q)}
\label{ds}.
\end{equation}
Then the sigma term in the derivative of Eq.~(\ref{pcorr0}) with respect to $G$, at fixed $M$ (since $M$ is determined by extremizing the pressure) is simply (\ref{Dint}) -- namely the right side of Eq. (\ref{PG}), with (\ref{ds})) -- while the first term is just the derivative of the first term on the right side of the mean field pressure, Eq.~(\ref{pmean}), at fixed $M$.

    Carrying out the Matsubara sum in Eq.~(\ref{pcorr0}), one has  
\begin{eqnarray}
& &p_{\rm corr} (T) =   \frac{i}{2} \int \frac{d^3q}{(2\pi )^3}  \int_{-\infty}^\infty \frac{d \omega} {2\pi} f_B (\omega)\nonumber \\ 
& & \times   \left\{ 3 \ln \left[ \frac{{\cal M}_\pi (\omega - i \epsilon,q)}{{\cal M}_\pi (\omega + i \epsilon,q)} \right]
+  \ln \left[ \frac{{\cal M}_\sigma (\omega - i \epsilon, q)}{{\cal M}_\sigma (\omega + i \epsilon,q)} \right] \right\}, \nonumber\\
\label{pcorr}
\end{eqnarray}
where
\begin{eqnarray}
{\cal M}_\nu (\omega , q)  \equiv  1-2G\Pi_\nu  (\omega, q) , 
\label{calm}
\end{eqnarray}
The arguments of the logarithms in Eq.~(\ref{pcorr}) can be interpreted as the phase shifts of the scattering of the $q\bar q$ pair in the time-like ($\omega > q$) region.   Contributions from the mesonic excitations arise from the zeros of ${\cal M}_{\pi/\sigma} $ in the complex $\omega$ plane.  \\

   Integrating by parts with respect to $\omega$ in (\ref{pcorr}), we derive the useful form,
   \begin{eqnarray}
p_{\rm corr} (T) & =  & - \int \frac{d^3 q}{(2\pi )^3} \int_0^\infty d \omega \nonumber 
\left( \frac{\omega}{2}  +  T \ln  ( 1 -  e^{- \beta \omega} ) \right) \\
& &\times \left[ 3 \rho_\pi (\omega, q; T) + \rho_\sigma (\omega, q; T) \right],
\label{pcorr2}
\end{eqnarray}
where we introduce the spectral weights
\begin{equation}
\rho (\omega, q; T)  =  \frac{1}{2\pi i} \left[ \frac{1}{{\cal M}_-}  \frac{\partial {\cal M}_-}{\partial \omega} 
- \frac{1}{{\cal M}_+ } \frac{\partial {\cal M}_+ }{\partial \omega}  \right] 
\label{spec1} 
\end{equation}
for $\pi$ and $\sigma$ mesons, with
${\cal M}_\pm (\omega, q) \equiv {\cal M} ( \omega \pm i \epsilon, q) $.
The spectral weights are real and can be written in terms of the real and imaginary parts of  $ {\cal M} ( \omega \pm i \epsilon, q) = {\cal M}_1 (\omega, q ) \pm i {\cal M}_2 (\omega, q ) $ as 
\begin{equation}
\rho (\omega, q; T)  =  \frac{1}{\pi} \frac{ {\cal M}_2 \partial {\cal M}_1 /\partial \omega  -{\cal M}_1 \partial  {\cal M}_2/\partial \omega }{{\cal M}_1 (\omega , q)^2  + {\cal M}_2 (\omega, q)^2} .
\label{spec2} 
\end{equation}
The $\omega/2$ term in (\ref{pcorr2})  is the vacuum pressure, whose divergent part is removed by renormalization at $T=0$,
leaving a finite contribution to the pressure at finite temperature.  

  We differentiate the correlation pressure (\ref{pcorr2}) with respect to $T$ and use the 
relation $\partial\left(T \ln  ( 1 -  e^{- \beta \omega} )\right){\partial T}
 =   \sigma_B (\omega)$, to find the entropy,
\begin{eqnarray}
s_{\rm corr} & = & - \int \frac{d^3 q}{(2\pi )^3} \int_0^\infty d \omega
\sigma_B (\omega)
\left[ 3 \rho_\pi (\omega, q; T) + \right. \nonumber  \\ 
&&  \qquad +  \left.  \rho_\sigma (\omega, q; T) \right]  
 + \Delta s_{\rm corr},
\label{scorr}
\end{eqnarray}
where
\begin{eqnarray}
\Delta s_{\rm corr} = - \int \frac{d^3 q}{(2\pi )^3} \int_0^\infty d \omega
\left( \frac{\omega}{2}  +  T \ln  ( 1 -  e^{- \beta \omega} ) \right) \nonumber \\
\times \left[ 3 \frac{\partial }{\partial T} \rho_\pi (\omega, q; T) + \frac{\partial }{\partial T} \rho_\sigma (\omega, q; T) \right]. 
\end{eqnarray}

The values of the quark mass $M$ and the Polyakov loop parameter $\Phi$ contained in $\Pi_{\pi}$ and $\Pi_\sigma$  minimize the free energy, and therefore, in calculating the entropy from the mesonic correlation pressure, both $M$ and $\Phi$ can be held fixed.  The temperature dependence of $p_{\rm corr}$ is contained in $f_B (\omega)$ as well as in the quenched quark distribution functions $f_\Phi (E_p)$ in the quark bubbles, $\Pi_{\pi}$ and $\Pi_\sigma$. 

 To proceed with the calculation of the correlation entropy, we write the explicit forms of the quark polarizations:
\begin{eqnarray}
& & {\rm Tr} \left[ \Gamma_\nu S_i (\varepsilon + \omega, p + q) \Gamma_\nu S_i (\varepsilon, p) \right] 
\nonumber \\
& & \quad = \frac{N_\nu}{\left[ (\varepsilon + \omega + i \phi_i T)^2 - E_{p+q}^2 \right]
\left[ (\varepsilon + i \phi_i T)^2 - E_p^2 \right] }
\nonumber \\
\label{pol}
\end{eqnarray}
where
\begin{eqnarray}
N_\pi  & = & 
{\rm Tr} \left\{ i \gamma_5 \left[ (\varepsilon + \omega + i \phi T) \gamma_0 + ( \bp + \bq) \cdot {\boldsymbol \gamma} + M \right] \right.
\nonumber \\
& & \qquad \qquad \times i \gamma_5 \left. \left[ (\varepsilon + i \phi T) \gamma_0 + \bp \cdot {\boldsymbol \gamma} + M \right] 
\right\}
\nonumber \\
& = & 
2\left( \left[ (\varepsilon + \omega + i \phi_i T)^2 - E_{p+q}^2 \right] 
 \right.
\nonumber \\
& & \left.\quad +  \left[ (\varepsilon + i \phi_i T)^2 - E_p^2 \right]-  (\omega^2 - q^2) \right),
\end{eqnarray}
and
\begin{eqnarray}
N_\sigma & = & 
{\rm Tr} \left\{ \left[ (\varepsilon + \omega + i \phi T) \gamma_0 + ( \bp + \bq) \cdot {\boldsymbol \gamma} + M \right] \right.
\nonumber \\
& & \qquad \qquad \times \left. \left[ (\varepsilon + i \phi T) \gamma_0 + \bp \cdot {\boldsymbol \gamma} + M \right] 
\right\}
\nonumber \\
& = & 
2 \left(\left[ (\varepsilon + \omega + i \phi_i T)^2 - E_{p+q}^2 \right]+ (\varepsilon + i \phi_i T)^2\right.
\nonumber \\
& & \left. - E_p^2  -  (\omega^2 - q^2 - 4M^2 )\right). \label{Nnu}
\end{eqnarray}
Inserting (\ref{pol})-(\ref{Nnu}) into (\ref{polarization1}) and performing the $\varepsilon$ integral, we pick up the 
residues of the poles at $\varepsilon = \pm E_{p+q} - \omega - i \phi_i T$ and at $\varepsilon = \pm E_{p} - i \phi_i T$,
and find
 \beq
  \Pi_\pi (\omega, q ) = {\cal T} + (\omega^2 - q^2){\cal F} (\omega, q),
 \eeq  
and 
\beq
  \Pi_\sigma (\omega, q ) =  {\cal T} + (\omega^2 - q^2 - 4M^2){\cal F} (\omega, q),
\eeq
where the constant tadpole term ${\cal T}$ equals  $\langle {\bar q} q \rangle /M$.
 Also  
\begin{equation}
{\cal F} (\omega, q) = {\cal F}_{\rm scatt}  (\omega, q) + {\cal F}_{\rm pair} (\omega, q),
\end{equation}
with the quark particle-hole bubble,
 \begin{eqnarray}
 {\cal F}_{\rm scatt} ( \omega, q  ) 
&=&  \gamma_q \int \frac{d ^3 p}{(2\pi)^3} \frac{f_\Phi  (E_p) - f_\Phi (E_{p+q})}{2 E_p\,\, 2 E_{p+q}} \nonumber\\
&&\times \left( \frac{1}{\omega+ E_p - E_{p +q}} - \frac{1}{\omega - E_p + E_{p +q}} \right), \nonumber\\
\eeq
and the quark-antiquark bubble,
\beq
{\cal F}_{\rm pair} (\omega, q ) 
& = & \gamma_q \int \frac{d ^3 p}{(2\pi)^3} \frac{1 - f_\Phi (E_p) - f_\Phi (E_{p+q})}{2 E_p\,\, 2 E_{p+q}} \\
&\times &\left( \frac{1}{\omega + E_p + E_{p +q}} - \frac{1}{\omega - E_p - E_{p +q}} \right).
\nonumber \\
\end{eqnarray}
We see explicitly that the $\Pi_\nu$ and hence the ${\cal M}_\nu$ are even functions of $\omega$, 

 The function $ {\cal F}_{\rm scat} ( \omega - i \epsilon, q  )$  has a non-zero imaginary part in the space-like region
$\omega < q$, while $ {\cal F}_{\rm pair} ( \omega - i \epsilon, q  )$  has a non-zero imaginary part in the time-like region $ \sqrt{ q^2 + 4 M^2} < \omega$.   The effect of the Polyakov loop appears in the suppression of the continuum due to the suppression of the quark distribution functions;
the kinematical conditions for the location of the continua are unchanged by a uniform temporal color gauge field, since the complex chemical potentials for quark and holes or antiquarks generated by the uniform gauge field cancel each other due to the color neutrality of the pair. 
The quark continuum contributions to the correlation pressure and entropy are strongly suppressed in the confining phase, and the mesonic entropy is  essentially that of free mesons.

  In the chiral limit ($m_0=0$) the gap equation (\ref{gap}) implies that for $M \neq 0$ below $T_c$,  ${\cal T}  = 1/2G$,  so that $\cal M$ contains only the quark bubble terms, and the entropy from the free meson gas is readily isolated from the non-collective quark pair excitations.   To recover the free meson results in the chiral limit below $T_c$, we note that since
${\cal M}_\nu (\omega, q)$ factors into $ (- \omega^2 + \omega{_q\nu}^2) {\cal F}$,  we can extract the piece in $\rho (\omega, q)$,  
\begin{equation}
\rho_{\rm meson} (\omega, q) = \delta (\omega - \omega_q) - \delta (\omega + \omega_q),
\label{mspec}
\end{equation}
which with Eq.~(\ref{pcorr2}) leads immediately to Eq.~(\ref{pmeson2}).
While the upper edge of the scattering continuum at the light cone coincides with the location of the pion pole and the lower edge of pair excitation continuum with the $\sigma$ meson pole in the chiral limit, the meson poles decouple from the continuum owing to the factorization of ${\cal M}_\nu (\omega, q)$.
 
On the other hand, for $T > T_c$, the gap equation has only the solution $M = 0$, and rather
\begin{equation}
1-2G{\cal T}\simeq c (T - T_c ).
\end{equation} 
Furthermore, the gap in the time-like continuum vanishes above $T_c$ since $M=0$, closing the window between two continua.
There is no room where the collective mesonic excitations appear as isolated poles.

   More generally, with a non-vanishing bare quark mass, $m_0$, explicitly breaking the chiral symmetry, 
${\cal M} (\omega, q)$ does not simply factorize, even at low temperatures. 
However, since the gap equation always has solutions with non-vanishing $M$, there will be a gap in the continuum excitation spectrum, and an isolated meson pole can exist in such window.       
At temperatures below $T_c$, ${\cal M} (\omega, q)$ still contains an isolated zero, corresponding to mesons with spectra shifted from those in the chiral limit, in the region $q < \omega < \sqrt{q^2 + 4 M^2}$ where the imaginary part of ${\cal F} (\omega \pm i \epsilon, q)$ vanishes \footnote{This is the case for pions up to a certain melting temperature above $T_c$, while the $\sigma$ meson pole is absorbed into the pair excitation continuum; the signal of pole remains as a strong peak since the continuum is suppressed by the Polyakov loop in the confining phase.}.
In this case we can still extract the meson pressure and entropy from the $\omega$-integral over this region.

Suppose ${\cal M}_\nu (\omega, q)$ has a zero at $\omega = \omega_q$; owing to ${\cal M}_\nu (\omega, q)$ being even in $\omega$, there is also a zero at  $\omega = - \omega_q$.  (We do not need to assume that $\omega_q = \sqrt{q^2 + m^2}$ here). 
In the vicinity of each of these zeros, 
${\cal M} (\omega, q) = c_q (\omega^2 - \omega_q^2)$, where $c_q$ is a non-vanishing function of $q$.  
Then the square bracket in (\ref{pcorr2}) reduces in the vicinity of $\pm \omega_q$ to the $\delta$-function form (\ref{mspec}) 
which upon integration over $\omega$ in Eq.~(\ref{pcorr2})
indeed gives the meson pressure in the form (\ref{pmeson2}) with a modified spectrum $\omega_q$ obtained from ${\cal M} (\omega_q , q) =0$.  
The corresponding meson entropy approximately assumes the canonical form (\ref{smeson}), with a small correction $\Delta s_{\rm corr}$ originating from the temperature dependence of the meson spectrum $\rho (\omega_q, q )$;  this dependence arises via 
the temperature dependence of $\omega_q$ as well as the more explicit temperature dependence of the distribution function.

\begin{figure}[htbp]
\begin{center}
\includegraphics[clip,width=65mm, angle=270]{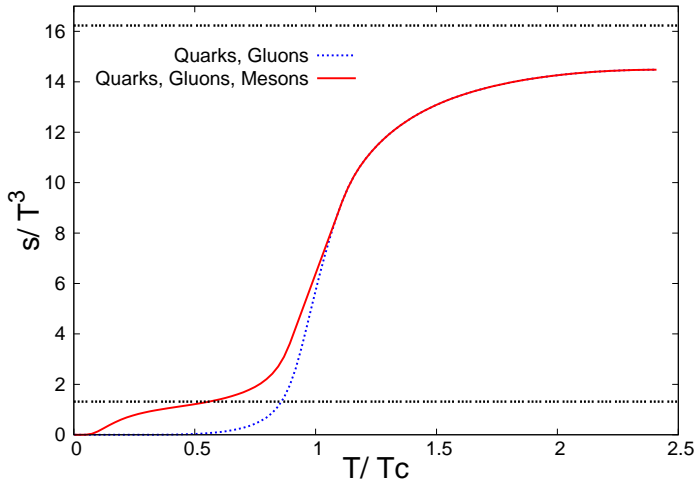}
\end{center}
\caption{
Temperature dependence of the entropy scaled by $T^3$:
the dotted line is the entropy carried by quarks and gluons in the mean field approximation, while the solid curve includes the entropy carried by the collective excitations.   The entropies of the massless pion gas and the gas of massless quarks and gluons are indicated by lower and higher horizontal dashed lines, respectively.  The temperature is scaled by $T_c$ taken here to be 230 MeV, of order that determined by the chiral susceptibility in the mean field approximation. }
\end{figure}

Additional contributions to the pressure and entropy arise from (non-collective) fluctuations of thermal quark quasiparticle excitations, due to scattering for $\omega <q$ and pair creation for $\omega >  \sqrt{q^2 + 4 M^2}$,  
as indicated by the appearance of a non-vanishing imaginary part of ${\cal F} (\omega \pm i \epsilon, q)$. 
As the temperature increases, isolated meson poles become absorbed into the continuum of quark-quasiparticle excitations, and no singularity appears in the pressure or entropy as a function of temperature.  
This non-collective correlation contains the remnants of the meson poles as resonance peaks in $\rho (\omega, q)$. 
As found numerically, the contributions of these non-collective quark fluctuations to the pressure is actually small compared to the mean field pressure of the quark quasiparticles \cite{Yamazaki:2012ux}.
In the chiral limit,  coupling between mesonic modes, both pions and sigma mesons,  and quark-pair excitations turns on suddenly at the critical temperature as the temperature is raised from below, causing a singularity in the derivative of entropy density with respect to the temperature, and changing the transition to second order, as in the mean field approximation.

\section{summary and conclusions}
We have examined here how the entropy in the PNJL model changes through the quark-hadron phase transition.
In  mean field, the entropy of the system does not take the simple form expected in a collection of quark-quasiparticle excitations or "quark triads" due to the color correlations induced by the Polyakov loop which suppress the quark entropy in the low temperature confining regime.  In addition to the quark-quasiparticle mean field entropy, the total entropy includes a contribution from mesonic correlations, which 
we have computed in the Gaussian approximation to the fluctuation of the auxiliary mesonic fields coupled to the quark-quasiparticle pair excitations, as well as in the more standard random phase approximation.
As the collective mesonic degrees of freedom melt with increasing temperature, the entropy they carry is transferred to the correlated pair excitations, in the background of increasing quark-quasiparticle mean-field entropy. 
Hence in this model the entropy changes continuously as the degrees of freedom change from those of hadrons to quarks and gluons,  with the transition second order in the chiral limit.
 
The analysis here is limited to zero chemical potential.  An interesting related question at non-zero chemical potential is how, as low density ordinary nuclear matter turns into high density quark matter  in a color superconducting phase \cite{SW99, YTHB07},  the carriers of baryon number change. We shall address this question in a future publication.  \\

 \noindent{\bf Acknowledgement} 

We thank Professor Tetsuo Hatsuda and Professor Kenji Fukushima for useful discussions.
KY would like to thank Professor Veronique Bernard and Professor Brigitte Hiller for informative discussions. 
KY's work has been supported by the University of Tokyo Grants for Ph.D. Research, 
Research Assistant for Creation of the Research Core in Physics, 
School of Science Grants for Ph.D. Students, and JSPS research fellowships for Young Scientists.   
TM's work is supported in part by the Grant-in-Aid \# 25400247 of MEXT, Japan.  
GB's work is supported in part by U.S. NSF Grant PHY13-05891.  In addition GB thanks the RIKEN iTHES project
for partial support during the writing of this paper.




\end{document}